\documentclass[11pt,twoside]{article}
\usepackage{FUSE2004}
\usepackage{natbib}

\usepackage{epsf}
\usepackage{psfig}
\usepackage{lscape}

\begin{document}
\title{P {\sc V} and the Mass Loss Discrepancy in O Stars}
\author{D.\ Massa}
\affil{SGT, Inc. \\Code 681, NASA's GSFC, Greenbelt, MD 20901, USA}
\author{A.W.\ Fullerton}
\affil{University of Victoria \& Johns Hopkins University\\3400 N. Charles St., 
Baltimore, MD 21218, USA}
\author{R.K.\ Prinja}
\affil{Department of Physics \& Astronomy, UCL\\Gower Street, 
London WC1E 6BT, UK}

\begin{abstract}
Building upon a previous analysis of P~{\sc v} wind lines in LMC O stars,  
we analyze the P~{\sc v} wind lines in a sample of Galactic O stars 
which have empirical mass loss rates determined from either their radio 
fluxes or H$\alpha$\/ profiles.  Since the wind analysis provides a 
measure of $\dot{M} q$ where $q$ is the ionization fraction of the ion, 
we determine $q($P~{\sc v}$)$ observationally.  In spite of model 
predictions that $q \sim 1$ for mid-O stars, we find $q($P~{\sc v}$) 
\leq 0.15$ throughout the O stars.  We discuss the origin of this 
discrepancy.
\end{abstract}

\section{Measures of $\dot{M}$}

Three approaches are normally used to determine stellar mass loss 
rates, $\dot{M}$s.  All assume that the wind is homogeneous and 
spherically symmetric (SS) with a single, monotonic velocity law, and 
all {\em should} agree.  The three approaches are:\\
\noindent {\it 1. Continuum excess from free-free emission.}  This 
samples the outer wind (the exact radius depends on wavelength), where it 
becomes optically thick to free-free emission.  It is only detectable 
for massive winds in nearby stars.  The radio wavelengths are considered 
``cleanest'', because, in contrast to the IR/FIR, massive winds become 
optically thick at large radii ($\geq 10 R_*$), where $v = v_\infty$,
a constant.
This makes $\rho_{wind}\sim \dot{M}/(r^2 v_\infty)$, independent of $v(r)$.  
Furthermore, no photospheric correction is needed.  However, the radio 
flux can be non-thermal, so observations at multiple wavelengths are 
required to determine the spectral index of the emission.\\
\noindent {\it 2. H$\alpha$\/ emission.}  This samples the inner wind and 
is easily observed.  For massive winds, H$\alpha$\/ emission is related to 
$\dot{M}$.  However, the exact form of the observed H$\alpha$\/ profile 
depends upon the $N=3$ departure coefficient for H in the wind and this, 
in turn, depends upon: the photospheric radiation field; the diffuse 
radiation field of the wind; and the wind velocity law in the 
acceleration region.  The shape of the ``photospheric'' H$\alpha$\/ 
profile is also required, and the observed $W_\lambda (H\alpha)$ can be 
strongly variable. Nevertheless, relatively sophisticated models for 
H$\alpha$\/ formation exist (e.g., Repolust et al.\ 2004), and can
provide reasonable agreement between available radio and H$\alpha$\/ 
$\dot{M}$s when H$\alpha$ emission is strong.  

\noindent {\it 3. UV resonance lines.}  These sample the entire wind.  
Their shapes are determined by the radial optical depth of the wind, 
$\tau_{rad} \sim \dot{M} q_i A_E$, where $A_E$ and $q_i$ are the abundance 
of element $E$ and its ionization fraction for stage $i$.  However 
observations of a dominant ion ($q_i \sim1$), of known abundance are 
required to estimate $\dot{M}$\/ directly, and the wind lines of abundant, 
dominant ions are saturated in winds massive enough to be detected in the 
radio or to have reliable H$\alpha$\/ $\dot{M}$s.

\section{{\it FUSE}\/ and P V}

{\it FUSE}\/ gives access to P~{\sc v}~$\lambda\lambda 1118, 1128$.  
P~{\sc v} is a surrogate for C~{\sc iv} (Massa et al.\ 2003) and  
$q_i \sim 1$ is expected for both ions in mid-O star winds.  Unlike $A_C$, 
$A_P \simeq Const$ over the life of an O star.  Furthermore, for scaled 
solar abundances, $\tau_{rad}($C~{\sc iv}$)/\tau_{rad}($P~{\sc v}$) = 661$, 
so to {\it detect}\/ P~{\sc v}, $\tau_{rad}($C~{\sc iv}$) \geq 50$ and 
saturated -- as is the case for stars with detectable radio fluxes and 
strong H$\alpha$ emission.

\medskip
\noindent {\bf P V in LMC O Stars:} 
The first large scale {\it FUSE}\/ study of P~{\sc v} was by Massa et al.\ 
(2003).  They performed SEI (Lamers et al.\ 1987) fits to P{~\sc v} wind 
lines, determining $\tau_{rad}$ for 25 LMC O stars.  They then used 
$\dot{M}$s predicted by the Vink et al.\ (2001) theory and $A_{P} = 
0.50\times$solar to find that $q(P\; V)$ peaked between 45--50~kK, as 
expected, but with a peak value $\leq 0.15$ (see Fig.\ 1).  This result 
implies a factor of 7 or more discrepancy between the expected and 
observed $\dot{M}$s.  There are three possible explanations for this 
result: \vspace{.05in}\\
\noindent {\it 1.}\/ The LMC P abundance scales differently from other 
  elements, \\
{\it 2.}\/ The theoretical $\dot{M}$s are incorrect for the LMC, or \\
{\it 3.}\/ The winds are not homogeneous and SS, but strongly 
clumped or structured.\\ 

\medskip
\noindent {\bf P V in Galactic O Stars (preliminary results):}
The Galactic $A_P$ is well determined (e.g., Catanzaro et al.\ 
2003), so abundance is not an issue for the Galaxy.  We are currently 
analyzing P~{\sc v} in {\it Copernicus}, {\it Orfeus} and {\it FUSE}\/ 
data for stars with radio and/or H$\alpha$\/ $\dot{M}$\/ estimates.  This 
eliminates the need for model $\dot{M}$s to derive $q$ for P~{\sc v}, and 
determines how well different mass loss rate indicators agree.  So far, we 
have analyzed 30 stars (11 more will be observed by {\it FUSE}), and the 
results are shown in Figure ~2.  These results are effectively identical 
to those from the LMC (Fig.~1), implying the same conclusions.

\section{Conclusions}
\begin{itemize}
\vspace{-0.1in}
  \item An erroneous P abundance is not the cause of the small LMC 
    $q(P\;V)$s.
\vspace{-0.1in}
  \item Strong clumping/porosity must be the root of the problem. 
\vspace{-0.1in}
  \item  Large scale clumping can also strongly affect the 
     radio and H$\alpha$\/ $\dot{M}$s, so these should be re-evaluated.
\vspace{-0.1in}
  \item \underline{The good news:} the different measures of $\dot{M}$\/ 
  are sensitive to different aspects of the wind flow, so bringing them 
  all into agreement (together 
  with the X-ray and O~{\sc vi} wind line observations) will provide 
  powerful constraints on models of how the winds are structured.
\end{itemize}


\begin{figure}[!ht]
\plotfiddle{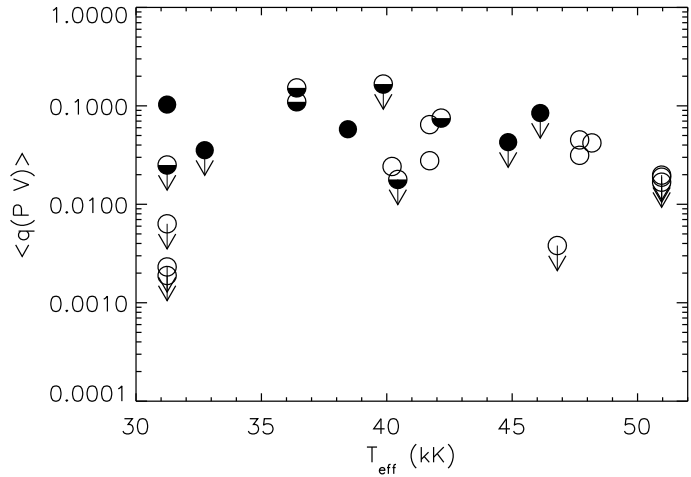}{2.4in}{0}{55}{55}{-170}{-210}
\caption{Mean P~{\sc v} ionization fractions as a function of temperature 
for LMC O stars analyzed by Massa et al.\ (2003).  Open, half filled and 
filled symbols denote stellar luminosities in the ranges $\log L/L_{\odot} 
> 6.0$, $6.0 \geq \log L/L_{\odot} > 5.6$ and $5.6 \geq \log L/L_{\odot}$, 
respectively.}
\end{figure}

\begin{figure}[!ht]
\plotfiddle{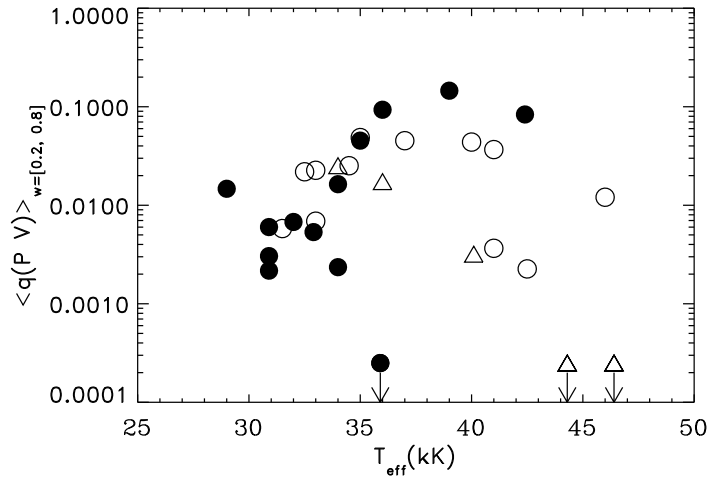}{2.25in}{0}{55}{55}{-170}{-210}
\caption{Mean P~{\sc v} ionization fractions versus temperature for 
Galactic O stars with radio or H$\alpha$\/ $\dot{M}$s.  Filled symbols: 
radio $\dot{M}$s, open circles: H$\alpha$\/ $\dot{M}$s from Repolust et 
al.\ (2004), open triangles: H$\alpha$\/ $\dot{M}$s from Lamers \& 
Leitherer (1993).}
\end{figure}

\end{document}